# Smart safety watch for elderly people and pregnant women

Balachandra D S[1], Maithreyee M S[2], Saipavan B M[3], Shashank S[4], Dr. P Devaki[5], Ms. Ashwini M[6]

[1]**Student, Department of Information Science and Engineering, The National Institute of Engineering, Mysore, Karnataka, India.**
[1*]`Balachandradevarangadi@gmail.com`

[2]**Student, Department of Information Science and Engineering, The National Institute of Engineering, Mysore, Karnataka, India.**
[2*]`mysoresampathmaithreyee@gmail.com`

[3]**Student, Department of Information Science and Engineering, The National Institute of Engineering, Mysore, Karnataka, India.**
[3*]`saipavanbm@gmail.com`

[4]**Student, Department of Information Science and Engineering, The National Institute of Engineering, Mysore, Karnataka, India.**
[4*]`shashanks33@outlook.com`

[5]**Professor, Department of Information Science and Engineering, The National Institute of Engineering, Mysore, Karnataka, India.**
[5*]`devaki@nie.ac.in`

[6]**Assistant Professor, Department of Information Science and Engineering, The National Institute of Engineering, Mysore, Karnataka, India.**
[6*]`ashwinim@nie.ac.in`

**Abstract.** Falls represent one of the most detrimental occurrences for the elderly. Given the continually increasing ageing demographic, there is a pressing demand for advancing fall detection systems. The swift progress in sensor networks and the Internet of Things (IoT) has made human-computer interaction through sensor fusion an acknowledged and potent approach for tackling the issue of fall detection. Even IoT-enabled systems can deliver economical health monitoring solutions tailored to pregnant women within their daily environments. Recent research indicates that these remote health monitoring setups have the potential to enhance the well-being of both the mother and the infant throughout the pregnancy and postpartum phases. One more emerging advancement is the integration of 'panic buttons,' which are gaining popularity due to the escalating emphasis on safety. These buttons instantly transmit the user's real-time location to pre-designated emergency contacts when activated. Our solution focuses on the above three challenges we see every day. Fall detection for the elderly helps the elderly in case they fall and have nobody around for help. Sleep pattern sensing is helpful for pregnant women based on the SPO2 sensors integrated within our device. It is also bundled with heart rate monitoring. Our third solution focuses on a panic situation; upon pressing the determined buttons, a panic alert would be sent to the emergency contacts listed. The device also comes with a mobile app developed using Flutter that takes care of all the heavy processing rather than the device itself.

## 1. Introduction

The World Health Organization (WHO) says that fall is ranked second in the number of accidental and unintentional deaths after road accidents. Identifying a fall cannot shield an elderly person from harm; the most effective strategy entails preemptive prevention. Nevertheless, responding promptly and providing assistance post-incident can substantially alleviate adverse consequences. Sleep disruptions while pregnant have been linked to an increased likelihood of preterm birth, gestational hyperglycemia, and mood disorders. Pregnancy stress could affect both the infant and the dynamic between the mother and the infant. Developing IoT-powered smart wearable devices, particularly tailored for women and children, holds the potential to swiftly address critical situations and avert potentially distressing incidents. These devices would offer panic alerts, real-time location sharing, and tracking capabilities, enhancing safety measures and minimizing the risk of traumatic encounters. These features will be provided with a wearable and an app to interact with the watch, built using the Flutter framework [5].

## 2. Literature Study

Several researchers have worked extensively on wearable-based Fall detection systems, and most of their papers have been very instructive in achieving our prototype/application. Some of them will be discussed in this section.

FalinWu **et al. (2015).** [1] and colleagues have developed a novel fall detection system that monitors the movements of the human body using motion-based sensors such as an accelerometer and gyroscope and uses an effective quaternion algorithm and informs the caregivers with the wearer's location. The system developed was low power consuming with an efficient algorithm.

Oktaf B Kharishma **et al. (2019).** [2] have tested the Neo 6m GPS and implemented it along with a use case to send a user SMS of location by the module via phone.

Lingmei ren **et al. (2019).** [3] have done a comprehensive study on various fall-detection based technologies, and two algorithmic approaches are discussed.

Based on existing literature, our proposed paper will showcase an application and prototype of a smart safety wearable with combined features of location tracking, fall detection and emergency alerts.

### 3. Methodology

#### 3.1 System Design

Flutter and Firebase are being used to create an app that will interact with the watch. The accelerometer, pulse oximeter, and GPS sensor are connected to Arduino Uno, which will get the readings. These will be transmitted to the app for computation using a Bluetooth module, and a display module is used on the watch to show alerts.

#### 3.1.1 Hardware design

**Arduino UNO:** The Arduino UNO is a microcontroller board developed by Arduino, leveraging the ATmega328P chip. It encompasses 14 digital input/output pins, six analog inputs, a 16 MHz ceramic resonator, a USB connection, a power jack, an ICSP header, and a reset button. As the central microcontroller, the Arduino UNO Rev3 will accommodate all sensors and the display within its framework [6].

**Pulse Oximeter & Heart Rate Sensor:** The MAX30102 [7] represents a comprehensive sensor solution for monitoring pulse oximetry and heart rate in an integrated manner Fig 1. It combines a pair of LEDs, a photodetector, finely tuned optics, and analog signal processing with minimal noise to effectively capture pulse oximetry and heart rate signals.

**OLED 12C Display:** A 0.96" OLED Display is used [13], connected to the Arduino using four wires. Two of which are for power and the other 2 for data. The data connection is I2C (Inter-Integrated Circuit) as shown in Fig 2.

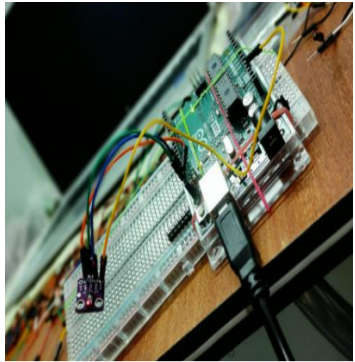
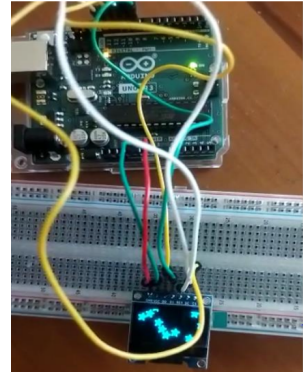

Fig. 1. MAX30100 sensor            Fig. 2. 0.96" OLED Display

**Bluetooth Module:** The HC-05 [8] Bluetooth Module employs a Serial Port Protocol module to facilitate a seamless setup for wireless serial connections. This module is crucial in establishing connections between watch and mobile applications. Additionally, it transmits data to the display and monitors oxygen levels and heart rate. Since multiple devices utilize this Bluetooth module, it has been programmed to function in a Master/Slave configuration Fig 3.

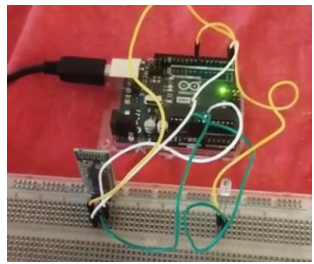
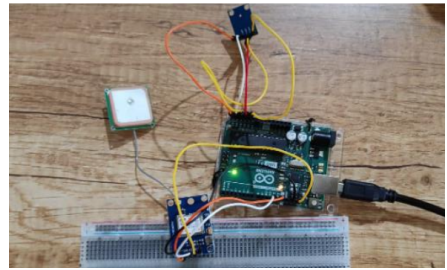

Fig. 3. HC-05 Bluetooth sensor            Fig. 4. NEO-6M GPS

Apart from these, a NEO-6M [2] GPS module Fig 4 which is used for tracking and getting current location, an ADXL335 3-axis accelerometer [1], which is used for fall detection, 9v Battery snap connectors used as a power supply for our Arduino boards, breadboard, 6*6*5 tactile push buttons, jumper wires and 9v batteries are used.

### 3.1.2 Software Design

The app designing and prototyping were done using Figma, and the app was developed using the Flutter framework.

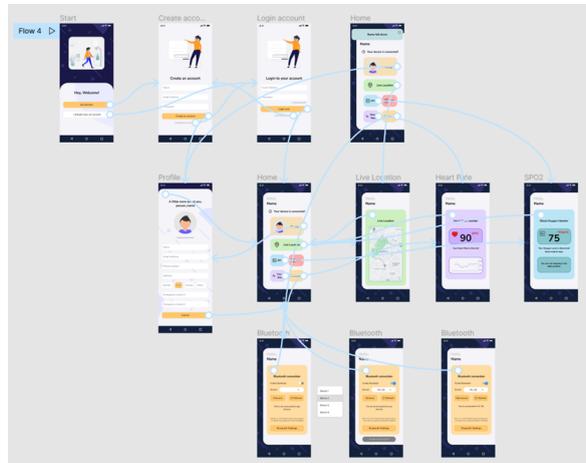

**Fig.5. Wireframe prototyping of the Mobile Application**

**Flutter:** Flutter is used as a cross-platform framework for developing apps on Android and iOS; this will be used to get the emergency contact details, get the readings from the watch using Bluetooth and track the location.

**Figma:** Figma serves as a tool for editing vector graphics, creating prototypes, and enabling the design of mobile apps; this is used as a prototyping tool for our app.

### 3.2 System Implementation

This section gives a concise portrayal of the workings of the current system. The system has been divided into two different modules: Hardware and Software.

**Scenarios**:

- When a fall is detected, especially in the case of the elderly, the fall detection algorithm tries to verify whether there is a fall. If it is a fall, the device will send a message to the listed emergency contacts.
- In the case of pregnant women [12], after clicking a button, the sensor starts accumulating the readings, which are then matched with the standard set of values. If values exceed the standard parameters, an alert will be sent to the user; if the user does not respond, a message is sent to the emergency contacts.
- In case of any emergency, the servers access the contacts to send an emergency message to the listed emergency contacts. If no contacts are listed, the device asks the user to select the emergency contacts available in the application.

### 3.2.1 Software Implementation

An app connects with the smartwatch and monitors spo2, heart rate, and live location, and sends mail and SMS to emergency contacts.

For login and registration, Firebase [9] authentication is used to maintain the auth state throughout the app, confirming email and resetting the password template. Fig 6 is also custom-designed, as shown below. While signing up, all emergency contact and pregnancy details are taken to use it down the lane further.

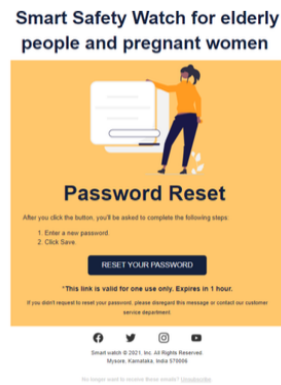
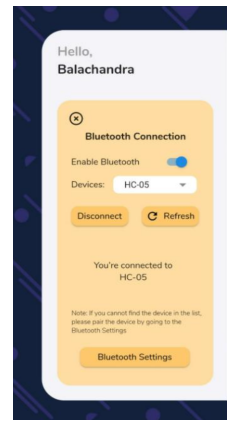

**Fig. 6. Email Template**      **Fig. 7. Bluetooth page**

#### 3.2.1.1 Connecting application to smartwatch

The app communicates with the smart watch using Bluetooth Fig. 8 where the watch is selected, so the Flutter Bluetooth serial package Fig. 7 is used to provide base functionality to select a Bluetooth device and get connected to it; this package is also used to maintain the connection after pairing to a smartwatch and transmit data to and fro watch.

#### 3.2.1.2 Making Application run in the background

The app is kept running in the background as the smartwatch syncs up regularly with the spo2 and heart rate readings; also, when a panic alert is triggered or a fall is detected, the watch will transmit it to the app, sending emails and SMS. A flutter background package is used for these features to work, forcing the app to run in the foreground and background and ensuring the task manager does not remove it.

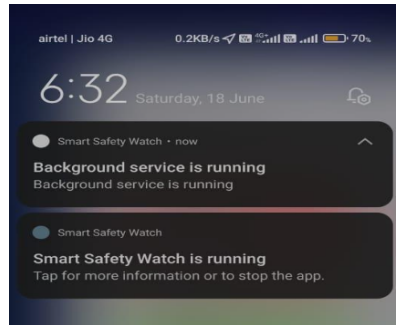

**Fig. 8. App running in the background**

### 3.2.1.3 Collection of Spo2 and heart rate readings

On a regular interval, when spo2 is switched on, the smartwatch scans and updates the value to watch, watch based on user profile and term of pregnancy, and will detect if the recorded spo2 value is standard or not when Bluetooth receives the data using provider, spo2 page is populated, and it is calculated if it is under normal range or not and if sleep position is proper. The same goes for heart rate readings.

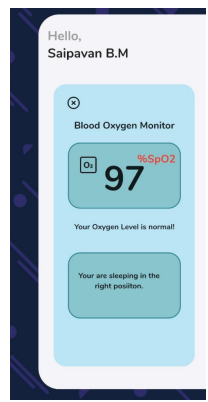 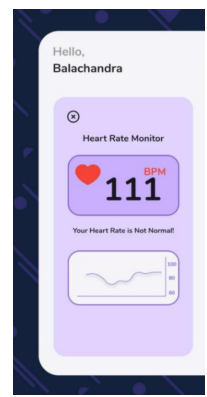

**Fig. 9. Spo2 page**            **Fig. 10. Heart Rate page**

#### 3.2.1.4 Location page for live tracking

Google Maps API [10] is used to add a map page to the app. The current location of the watch wearer is updated in real-time, and it is sent to emergency contacts mail and SMS when panic or fall is triggered. Geocoding is used to convert latitude and longitude to a specific address. The background location package is used to get location even when the app is in the background or foreground.

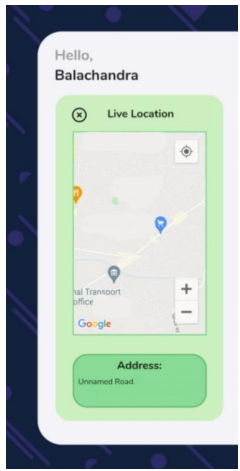

**Fig. 11. Location page          Fig. 12. Calibrating the accelerometer
sensor**

#### 3.2.1.5 Fall and panic alert functionality

When the watch transmits the data and the app detects the keyword for fall and SOS, SMS service and telephony package will be used to send messages to an emergency contact. It uses the http package to send a post API call to an email service called EmailJS, where an email template is stored and on receiving an API call from the app, it sends mail to the listed emergency mail ID.

### 3.2.2 Hardware Implementation

The prototype uses two Arduino UNOs, a GPS sensor, an Oximeter and heart rate sensor, a Bluetooth module, an OLED Display and an accelerometer.

The device is powered by two 9V power supply batteries, each of which powers a single Arduino UNO board and its sensors. One board has been integrated with the Bluetooth and OLED display. The other board has been integrated with SPO2, Accelerometer, GPS and Buttons. Two buttons are being used for Panic alerts and for prompt confirmation for Fall detection and sleeping position detection.

```
int roll = ( ( (atan2(yg,zg) * 180) / 3.14 ) + 180 );
int pitch = ( ( (atan2(zg,xg) * 180) / 3.14 ) + 180 );
int yaw = ( ( (atan2(xg,yg) * 180) / 3.14 ) + 180 );
```

**Fig. 13. Calculating the Roll, Pitch and Yaw used to detect falls.**

The ADXL335 accelerometer sensor is used to detect the orientation of a person and the acceleration during movement. It comprises three axes - X, Y and Z. Each of the axes can calculate the acceleration in that direction individually. When the accelerations of the three axes are used together, metrics such as Roll, Pitch and Yaw are identified, which are used to check the change in the angle of orientations.

The accelerometer is initially calibrated by placing it on a flat surface, and the calibrated values are later used for computation. The root mean square acceleration is calculated using all three acceleration values. A threshold for the RMS value is set; upon crossing this threshold value, it can be said that the person has fallen.

Displaying data [11] on the screen and sending data to the device using Bluetooth:

*Serial.readBytes(mystr,14);*
*display.println(mystr);*
*Serial.println(mystr);*
*display.display();*

*mystr.toCharArray(char_array, str_len);*
*EEBlue.write(char_array,str_len);*

```
long irValue = particleSensor.getIR();
currentMillis = millis();
//Serial.println(currentMillis);
if (checkForBeat(irValue) == true)
{
  //We sensed a beat!
  long delta = millis() - lastBeat;
  lastBeat = millis();

  beatsPerMinute = 60 / (delta / 1000.0);

  if (beatsPerMinute < 255 && beatsPerMinute > 20)
  {
    rates[rateSpot++] = (byte)beatsPerMinute; //Store this reading in the array
    rateSpot %= RATE_SIZE; //Wrap variable

    //Take average of readings
    beatAvg = 0;
    for (byte x = 0 ; x < RATE_SIZE ; x++)
      beatAvg += rates[x];
    beatAvg /= RATE_SIZE;
  }
}
```

**Fig. 14. Obtaining and measuring the SPO2 and Heart Rate readings**

Heartbeat is calculated based on the amount of light absorbed, and it uses a particle sensor library function to collect the readings. The function collects 100 sample readings, averages all the readings, and then displays the final Heart rate.

The SPO2 Module obtains the readings using photodetectors and bright LED Lights. This sensor calculates a person's oxygen level by the amount of red light and IR rays passing through the person's finger. The deoxygenated light absorbs red light more than oxygenated blood absorbs IR light. The ratio of these readings used for the blood's oxygen level will be calculated.

For panic alerts, buttons are attached to the Arduino when pressed, a SOS signal is sent to the app, and the app then triggers SMS and Mail to emergency contacts. When the button is pressed, serial write happens, and it is communicated to the second Arduino board, where it transmits using the Bluetooth module to the application.

```
bool currentState=digitalRead(BUTTON_PIN);
if(currentState==pressed)
{
  char mysostr[5] = "SOS\n";
  Serial.write(mysostr,5);
  delay(500);
}
```

**Fig. 15. Sending a panic alert to the OLED screen**

### 3.3 Flowcharts

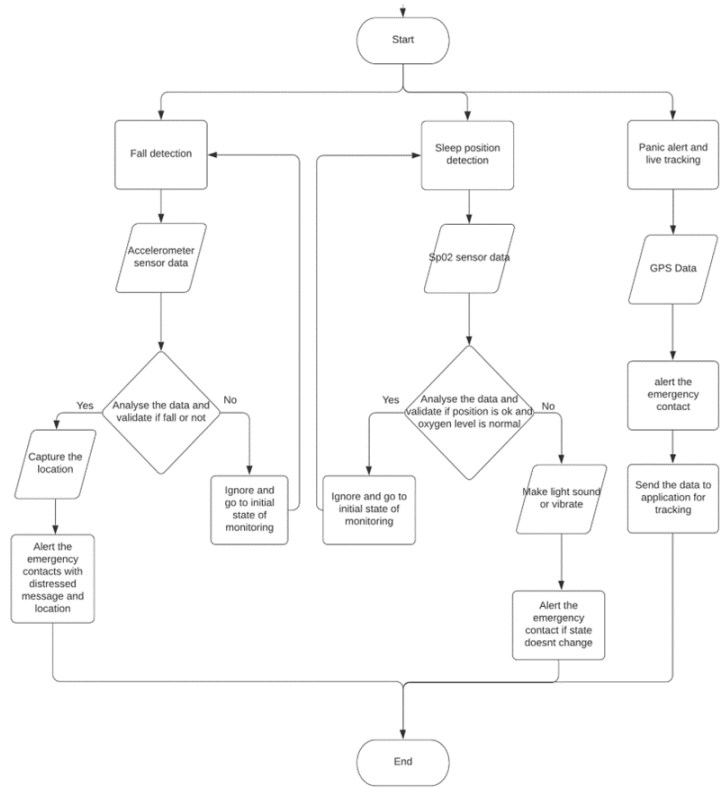

**Fig. 16. Detailed Design**

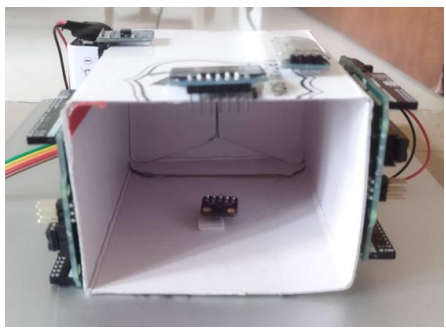

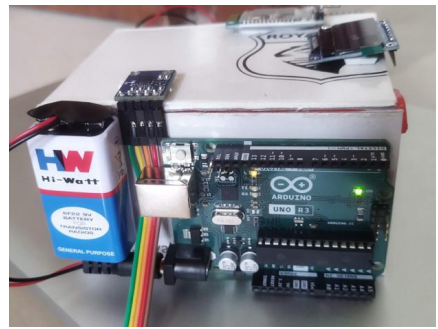

**Fig. 17. Inner view of spo2 and heart rate display**

**Fig. 18. Accelerometer and**

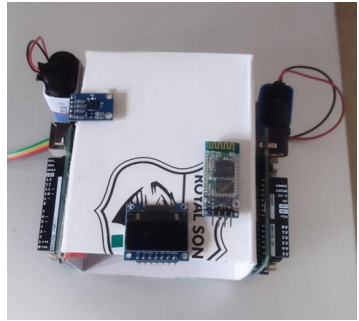 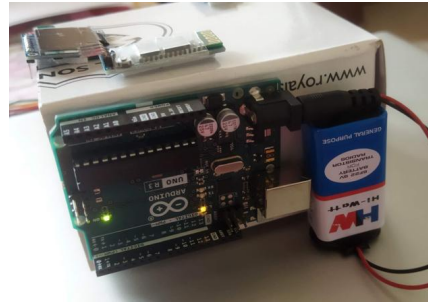

**Fig. 19. Watch prototype before connections**     **Fig. 20. Bluetooth module**

### 3.4 Issues and Challenges

Many issues were encountered initially, such as memory overload as a single Arduino was not enough to connect all the sensors, so we created a bridge between two Arduinos and sensors were divided between them. Then, there was an issue of arranging the setup so that someone could wear it and keep it as close to the watch as possible, which we overcame by using double-layer cardboard, which would be our scaffold for holding the Arduino boards and concealing the wires. The size is still an issue; this was just a prototype, which would be our proof of concept. Some disadvantages include the energy consumed, which will be more as various applications run simultaneously, draining the battery. A constant connection between the mobile application and the device must function correctly. Being a prototype, the current system is larger than a wearable device. GPS Connection is not as accurate as a three GPS setup would be more accurate, and this GPS may not give the correct location in an area that is covered by tall buildings.

### 4. Evaluation Metrics and Testing

**Test Cases:**

| S. No | Test Case | Expected Result | Actual Result | Result State |
|---|---|---|---|---|
|  |  |  |  |  |
| 1. | Does the Bluetooth HC05 module work and connect with the app? | The sensor blinks fast while not connected and slow when connected to the app. | The sensor blinks fast while not connected, and slow when connected to the app. | Correctly predicted |
| 2. | Does the GPS module work, and is the location transmitted to the app? | The module blinks and location values are obtained and transmitted to the app. | The module blinks and location values are obtained and transmitted to the app. | Correctly predicted |
| 3. | Does the MAX30102 GY SPO2 sensor work and transmit to the app? | On connecting, a constant red light is visible and takes input and transmits it to the app. | On connecting, a constant red light is visible and takes input and transmits it to the app. | Correctly predicted |
| 4. | Does the ADXL 335 accelerometer work? | On connecting the module, the X, Y, and Z axis values' acceleration must be obtained. | On connecting the module, the X, Y, and Z axis values' acceleration must be obtained. | Correctly predicted |
| 5. | Does the display work? | The display on the connection shows blue text on the screen. | The display on the connection shows blue text on the screen. | Correctly predicted |

## 5. Conclusion

Since fall detection is a significant challenge in the healthcare domain, especially for elderly people, an IoT-based wearable fall detection system was built that'll help the elderly to get appropriate actions that are required in the case of a fall. Maternal health monitoring is essential to ensure the health and well-being of the mother and her child, as many health complications occur during pregnancy with a lifetime effect on their health.

**Results:**
- The detection of elderly falls is an example of the potential of autonomous health monitoring systems.
- The device will help constantly monitor the spO2 levels both in mother and child and send an alert if there are changes in the levels.
- This device also provides a live tracking mechanism that can be activated by clicking the button twice in case of any emergency, and this is sent via message to the emergency contacts listed.